\documentclass{sigchi-ext}
\usepackage[T1]{fontenc}
\usepackage{textcomp}
\usepackage[scaled=.92]{helvet} 
\usepackage{graphicx} 
\usepackage{balance}  
\usepackage{booktabs} 
\usepackage{ccicons}  
\usepackage{ragged2e} 
\usepackage{subcaption} 
\usepackage[nocompress]{cite}

\def\plaintitle{Algorithmic nudge to make better choices: Evaluating effectiveness of XAI frameworks to  reveal biases in algorithmic decision making to users} 
\def\emptyauthor{}
\def\plainkeywords{explainability, bias, transparency, intervention}

\title{Algorithmic nudge: Using XAI frameworks to design interventions}

\numberofauthors{6}
\author{%
  \alignauthor{%
    \textbf{Prerna Juneja}\\
    \affaddr{University of Washington} \\
    \affaddr{Seattle, WA,USA} \\
    \email{prerna79@uw.edu} } \alignauthor{%
    \textbf{Tanushree Mitra}\\
    \affaddr{University of Washington}\\
    \affaddr{Seattle, WA,USA}\\
    \email{tmitra@uw.edu} }
    }

\definecolor{linkColor}{RGB}{6,125,233}
\hypersetup{%
  pdftitle={\plaintitle},
  pdfauthor={\emptyauthor},
  pdfkeywords={\plainkeywords},
  bookmarksnumbered,
  pdfstartview={FitH},
  colorlinks,
  citecolor=black,
  filecolor=black,
  linkcolor=black,
  urlcolor=linkColor,
  breaklinks=true,
}

\begin{document}


\CopyrightYear{2020}
\setcopyright{rightsretained}
\conferenceinfo{ACM CHI Workshop on Operationalizing Human-Centered Perspectives in
Explainable AI}{May 8–9, 2021, Virtual}
\isbn{}
\doi{}
\copyrightinfo{\acmcopyright}

\maketitle

\RaggedRight{} 

\begin{abstract}
  In this position paper, we propose the use of existing XAI frameworks to design interventions in scenarios where algorithms expose users to problematic content (e.g. anti vaccine videos). Our intervention design includes \textit{facts} (to indicate algorithmic justification of what happened) accompanied with either \textit{fore warnings}  or \textit{counterfactual explanations}. While fore warnings indicate potential risks of an action to users, the counterfactual explanations will indicate what actions user should perform to change the algorithmic outcome. We envision the use of such interventions as `decision aids' to users which will help them make informed choices.
\end{abstract}

\keywords{\plainkeywords}

\ccsdesc[500]{Human-centered computing~Human computer interaction (HCI)}
\ccsdesc[100]{Human-centered computing~User studies}


AI algorithms play an important role in governing and shaping our lives. From recommending what websites to browse, what movies to watch, and what books to read, to informing decisions about defendants in the criminal justice system, these black-box algorithms play a crucial role in several low and high stake tasks. As the algorithmic systems become more pervasive, there has been a widespread concern about their role in amplifying or reinforcing various biases. Thus, researchers and scholars have pushed for making algorithms more accountable and transparent. This push has propelled the field of explainable AI (XAI) where the goal is to provide an explanation on how the machine learning algorithm reached a particular decision \cite{xai_goal}. The literature on XAI is vast (see \cite{inproceedings_lit_review} for a review). A few notable directions include explaining ML classifiers and multi-agent systems \cite{guidotti2018survey,rosenfeld2019explainability,alqaraawi2020evaluating,ANGELOV2020185}, designing guidelines for generating explanations \cite{doshi2017towards,kulesza2015principles,lakkaraju2016interpretable}, defining taxonomy of user needs for AI explainability \cite{liao2020questioning,lim2009assessing,lim2009and}, developing metrics and frameworks for evaluating explanations \cite{hoffman2018metrics,carvalho2019machine} and identifying various stakeholders of XAI \cite{preece2018stakeholders}.  Another line of scholarly research has also tested the effectiveness of the explanations used by online social media platforms. For example, investigation of Facebook's \textit{Why am I seeing this ad?} feature revealed that users often found the ad explanations misleading and incomplete \cite{andreou2018investigating}. There is also a burgeoning field of human-centered XAI where scholars draw from formal theories in HCI to inform the design of explanation interfaces \cite{miller2019explanation,hoffman2017explaining_1,hoffman2017explaining_2,klein2018explaining} and conduct empirical studies to 
examine how users interact with explanations \cite{cai2019effects,cheng2019explaining,hohman2019gamut,dodge2019explaining}. Existing literature reveals that XAI has the capability to not only explain the algorithmic decision making process but also provide a signal of how the algorithm will behave in future \cite{darpa}. 

In this position paper, we explore use of existing XAI frameworks in designing persuasive interventions in scenarios where user behaviour can lead to exposure to problematic content. In other words, we want to explore whether feedback from algorithms (designed using XAI frameworks) improve human decision making? 

\begin{figure*}
  \centering

      \centering
      \includegraphics[width=0.9\textwidth]{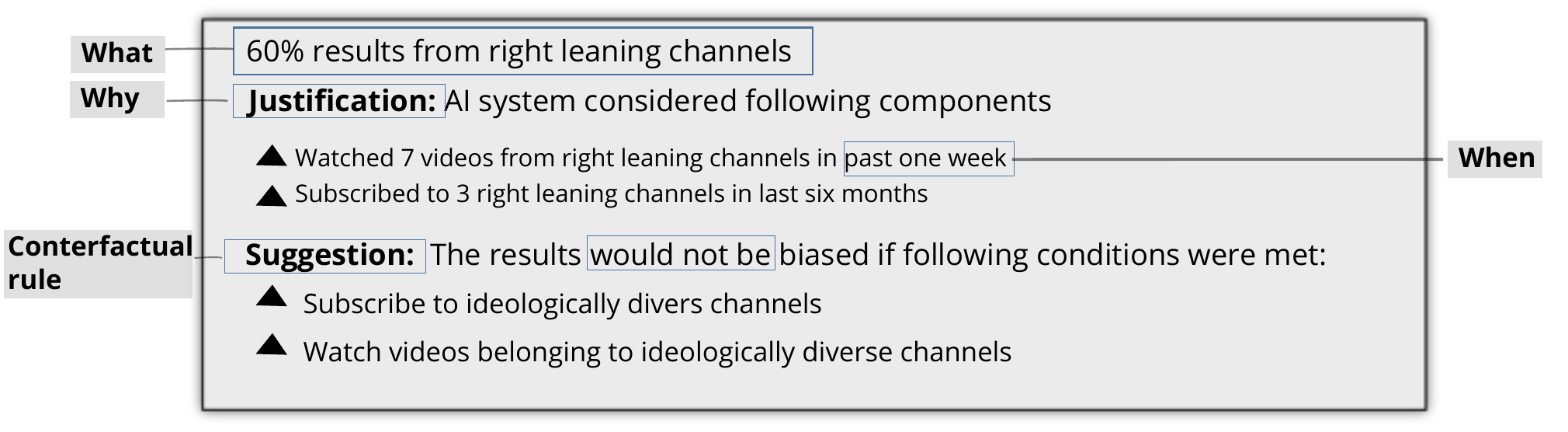}
      \caption{Scenario 1: When a user searches for a political query in YouTube and is presented with biased search results. Apart from algorithmic justification,  user is also presented with actions to change the algorithmic output in future.}
      \label{f1}
 \end{figure*}
\begin{figure*}
  \centering
      \centering
      \includegraphics[width=0.9\textwidth]{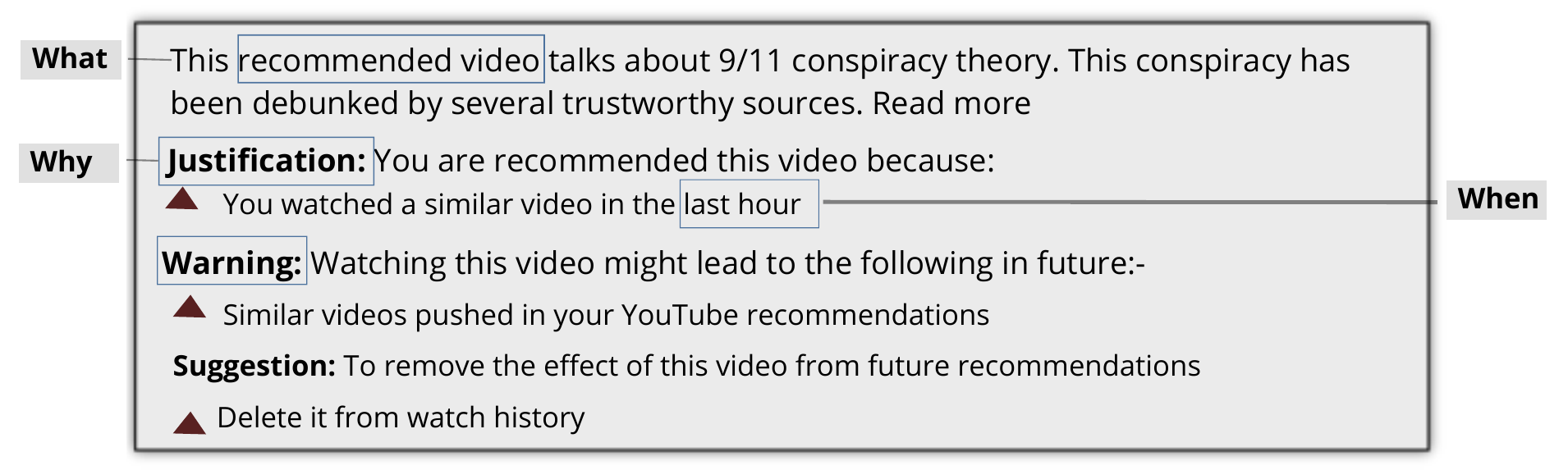}
      \caption{Scenario 2: When a video about 9/11 conspiracy theory appears in user's homepage. Here, apart from algorithmic justification, user is also warned about the consequences of watching this video in future.}
      \label{f2}
 \end{figure*}

We propose design using (1) facts (what happened and why) accompanied with forewarnings (what could happen) to convey the potential risks of an action in a comprehensible manner.  For example, consider a situation where a user searches for query ``election fraud proof'' on YouTube. The message should then forewarn users about the risk of their future video recommendations getting polluted from election misinformation content and ask them to re-think about their actions. (2) facts (what happened and why) accompanied with counterfactual explanations (what needs to change for another outcome to appear). This design would include message informing users why they are seeing a problematic content along with instructions of how could they sanitize their own content to not see that problematic content. For example, consider a situation when an anti-vaccine video appears in user's homepage. The message explains why such an occurrence occurred \textit{"This happened because..."} and remedies in a counterfactual tone \textit{"For this not to happen do........"}. In this particular example, counterfactual solutions could include suggesting users to delete the video from their search and watch history to remove its influence on future recommendations \cite{youtubehelp} or to  click on “not interested” to signal to YouTube that you are not interested in seeing this video. 

Scholars have argued that explanations are not always necessary or desirable and explaining everything in every situation \cite{deloitte} and can lead to information overload \cite{ehsan2021expanding}. Keeping this principle in mind, we propose our interventions in high stake scenarios where user actions have or could lead to more exposure of problematic content. We envision the use of such explanations as \textit{decision aids} to help users make better choices. It is important to note that defining problematic content is out of scope for this position paper. For the purpose of this proposal, we consider two examples of problematic content. First where algorithmic output has a partisan bias and second where users are presented with misinformation. We include partisan bias in this category because scholars have argued that \textit{selective exposure} to information from a specific ideology could lead to fragmented society \cite{babaei2018purple}.

We make use of two frameworks and theories to design our interventions. First we, use XAI design framework suggested by Ehsan et al based on the principles of Social Transparency that suggests use of design features reflecting the What, Why, Who, and When of user interactions with AI systems \cite{ehsan2021expanding}. In our design, `what' is conveyed by indicating a problematic behaviour and `when' is expressed by a timestamp, `why' is indicated in the algorithmic justification of `what'. `Who' is the user receiving these interventions, thus we do not explicitly mention that in the design. Next we make use of Fogg's behaviour change model (FGB) that has been used to design persuasive technologies \cite{10.1145/1541948.1541999}.  The model states that for a user to change behaviour, they must be (1) motivated, (2) have the ability to perform the change and should be triggered to perform the change \cite{10.1145/1541948.1541999}. Based on this model, we provide `explanations' as triggers, `bias indicators' as motivation to change behaviour and also `state the actions' that are required to change behaviour. We demonstrate the design of our XIA based interventions via few example scenarios in Figures \ref{f1} and \ref{f2}. We plan to test the effectiveness of such a design using user studies.

There are several open questions that our proposed design does not address. What are the various high-stake problematic scenarios that demand algorithmic interventions? How frequently should such interventions appear? Where should these interventions appear?  What granularity of algorithmic justification should appear in the design? What role will algorithm skepticism play in users acceptance or rejection of these interventions? We hope to discuss these questions during the workshop. 

\balance{} 

\bibliographystyle{SIGCHI-Reference-Format}
\bibliography{sample}

\end{document}